\documentclass[11pt,a4paper]{article}

\usepackage{pdfpages}
\RequirePackage{amsthm,amsmath,amsfonts,amssymb}
\RequirePackage[authoryear]{natbib}
\RequirePackage[colorlinks,citecolor=blue,urlcolor=blue]{hyperref}
\RequirePackage{graphicx}

\usepackage{subcaption}
\usepackage{wrapfig}
\usepackage{caption}
\usepackage{xfrac}
\usepackage{afterpage}
\usepackage[colorinlistoftodos]{todonotes}

\usepackage[textheight=23cm]{geometry}

\title{Some recent trends in embeddings of time series and dynamic networks}

\author{Dag Tj{\o}stheim$^{1,2}$ \and Martin Jullum$^2$ \and Anders L{\o}land$^2$}

\date{$^1$University of Bergen\\%
  $^2$Norwegian Computing Center
  }

\begin{document}

\maketitle

\begin{abstract}
 We give a review of some recent developments in embeddings of time series and dynamic networks. 
 We start out with traditional principal components and then look at extensions to dynamic factor models for time series. 
 Unlike principal components for time series, the literature on time-varying nonlinear embedding is rather sparse. 
 The most promising approaches in the literature is neural network based, and has recently performed well in forecasting competitions. 
 We also touch upon different forms of dynamics in topological data analysis. 
 The last part of the paper deals with embedding of dynamic networks where we believe there is a gap between available theory and the behavior of most real world networks.
 We illustrate our review with two simulated examples.
 Throughout the review, we highlight differences between the static and dynamic case, and point to several open problems in the dynamic case.
\end{abstract}

\section{Introduction \label{Intro}}

Traditional time series analysis handles equidistant observations in time, scalars or vectors. If it is a vector time series, the number of components is typically small or moderate, a possible exception being a panel of time series.

A characteristic feature of the Big Data revolution is that observations may be multivariate with literally millions of components. This has put restrictions on some of the traditional statistical methods and prompted the introduction of new ones, the latter often being treated in the machine learning literature. The large number of components have necessitated the search for new embeddings methods, where an embedding is assumed to keep original characteristic features of the data, but in a much lower-dimensional space, making further analysis easier and manageable. Principal component analysis is probably still the most used embedding method, but it cannot cope with data of ultra-high dimension due to the fact that then an ultra-dimensional eigenvalue problem is involved. Moreover, the data may be nonlinear in structure where the dependency relationships cannot be well modeled by variances and covariances as required by principal components analysis. Finally, ordinary principal components are not dynamic in nature, not taking care of auto dependencies in its construction, although that particular problem has been nearly fully addressed in the last two decades through the concepts of dynamic principal components and dynamic factor analysis. 

A number of nonlinear alternative embedding methods exist, and are surveyed in \cite{tjos:jull:lola:2022}. In that paper, we also reviewed topological data analysis (TDA), such an analysis being able to handle data with voids and cavities.

Another feature in current data analysis is that data often come in the form of
networks or graphs. The mathematical theory of graphs dates far back in time, but the statistical analysis of networks is of much more recent date. In fact there has been a tremendous growth in the literature on networks and their applications. Again we refer to \cite{tjos:jull:lola:2022} for a review of this development.

The survey in \cite{tjos:jull:lola:2022} (TJL in the sequel) is virtually completely limited to a static situation, and in fact an overwhelming part of the literature is restricted to this situation. But it is clear that in many cases the static assumption fails. Consider for example a network of bank customers. Such a network is changing in time: the strength of relationships between customers will in general change, and there may appear new customers joining the network and others leaving. This causes problems for instance in use of network methods in the detection of bank fraud such as money laundering \citep{jul:lol:hus:aan:lor:2020}. 

Another example of a time changing network occurs in imaging of changes in a brain network \citep{kucyi2017dynamic}. This could be caused by certain actions of a person or an animal participating in an experiment. In another medical experiment one may be interested in studying developments in time of certain topological patterns of the brain. 
An obvious third example is online social networks where `friends' and followers connect and interact with each other, enter and leave groups as times go by \citep{sarkar2005dynamic}.

How does this mesh with available modeling tools for analyzing such cases? Does there exist a statistical theory for time-varying embeddings and time-varying networks, and can it be applied to tasks as those mentioned above. The answer to this is a partial yes. Relevant methods are in the process of being developed, but many unsolved problems remain. This means that this area of research may be a treasure trove for time series analysts looking for new and inspiring problems. Much of the research so far is very recent and mostly found in the machine learning literature, often in the form of preprints. 
A main goal of this paper is to try, in a time-varying context,  to focus on possibilities for building a bridge between algorithmic machine learning and a more model based statistical approach. This will be done by advancing from TJL to a survey of  very recent and challenging problems in time series and dynamic networks embeddings.

In Section \ref{Principal} we start with the principal component embedding method in time series and the classical work of David Brillinger. Inspired by his work in the last couple of decades there have been important developments in dynamic principal component analysis and dynamic factor analysis. We also explain why it does not always succeed. Alternative time series embeddings are discussed in Section \ref{Nonlinear} by starting out with nonlinear statistical embedding methods such as for instance multidimensional scaling and local linear methodology. Time-varying topological data analysis is the topic of Section \ref{Dynamictopological}, and in Section \ref{Dynamic graphs} we treat temporary variation and time series in networks. In the paper we extend two illustrative examples from TJL to a situation where we look at robustness of a number of methods to certain changes in time. We should also stress that in reading this paper, even though an attempt has been made to make it self-contained,  it may be an advantage to have the general paper TJL at hand, where the fundamental concepts and definitions can be looked up in more detail.

\section{Principal components in time series \label{Principal}}

In general the principal component method is probably the most used method for reducing the dimension of data but still trying to retain the essential information. For iid (independent identically distributed) vector data this makes the principal component method to a yardstick against which other embedding methods may be compared. For time series  the situation is less clear.

As is well known, for $p$-dimensional observations the population principal components $V_j, j=1,\ldots,p$ are obtained by solving the eigenvalue problem 
\begin{equation}
\boldsymbol \Sigma V_j = \lambda_j V_j,
\label{2.1}
\end{equation}
where $\boldsymbol \Sigma$ is the $p \times p$ dimensional population covariance matrix of the data. Let the observations $X_i, i=1,\ldots,T$ have components $X_{ij}$, and let ${\bf X}$ be the matrix ${\bf X} = \{(X_{ij}-\bar{X}_j)\}$, with $\bar{X}_j = T^{-1}\sum_iX_{ij}$, then an estimate of $\boldsymbol \Sigma$ is obtained from $T^{-1}[{\bf X}^T{\bf X}]$, and the estimated principal components are obtained from
\begin{equation}
{\bf X}^T{\bf X}\hat{V}_j = \hat{\lambda}_j\hat{V}_j.
\label{2.2}
\end{equation}
%In TJL the two major weaknesses of the PCA (principal component analysis) were pointed out. 
There are two main weaknesses of the PCA (principal component analysis). The eigenvalue equations are based on second moments only, and if the dependencies of the data are not well described by second moments, as is not infrequently the case for financial time series, the value of principal components could be strongly reduced, and one may profit by using some of the nonlinear methods laid out in the next section (see also TJL). The second weakness of PCA is that for high dimensional data (large $p$) the eigenvalue problem may be cumbersome and time consuming to solve.

For time series there is another difficulty. The PCA method, as applied traditionally, neglects all dependence in time inherent in say auto and cross correlations in time. In fact there are many examples where auto
dependencies are simply ignored, for example in portfolio analysis of stock data by PCA, neglecting that stock data do not consist of iid data. The consequences of using ordinary PCA on time series data and possible associated pitfalls have recently been analyzed by \cite{zhan:tong:2021}. These authors derive asymptotic theory for eigenvalues and eigenvectors for the PCA eigenvalue equation \eqref{2.2}, under several forms for time series dependence. This problem has been considered earlier, notably by the recipient of this birthday volume, Masanobu Taniguchi, in \cite{tani:kris:1987}.

A more serious problem, though, is the structural difficulties that can emerge (or are likely to emerge) if auto dependence is neglected. For example a principal component that is obtained as one of the least significant in terms of a low eigenvalue, may actually be of central importance because it may happen to have a strong time dependence in terms of autocorrelation.

Is there  an alternative approach where this general problem is taken into account? The answer to this question is yes, and is constituted by the classical contribution of David Brillinger in \cite{bril:1969} and in Chapter 9 of his book \cite{bril:1975}. The devise used by Brillinger was to replace the covariance matrix $\boldsymbol \Sigma$ by the $p \times p$ dimensional spectral density matrix
\begin{equation}
{\bf {\bf f}(\omega}) = (2\pi)^{-1}\sum_{u = -\infty}^{\infty}{\boldsymbol \Sigma}_u\exp\{-i\omega u\},
\label{2.3}
\end{equation} 
where ${\boldsymbol \Sigma}_u$ is the cross covariance matrix at lag $u$. This expression includes all the second order dependence in time, and it gives a PCA decomposition in the frequency domain by solving the eigenvalue problem
$$
{\bf f}(\omega)V_j(\omega) = \lambda_j(\omega)V_j(\omega),
$$
resulting in principal components $V_j(\omega)$ depending on frequency. 
By taking the inverse Fourier transform of the eigenvectors of the spectral density matrix, the co-called dynamic principal components are the result. This procedure assumes that the time series is stationary, guaranteeing the existence of the spectral distribution and in addition that it is absolutely continuous so that the spectral density matrix exists. In \cite{pena:yoha:2016} the stationarity assumption is dropped. The general scheme and idea of Brillinger is kept, but \cite{pena:yoha:2016} allows for dynamic principal components that may not be linear combinations of the observations, and they may be based on a variety of loss functions, including robust ones. On the other hand this more general framework restricts the possibility of undertaking a rigorous statistical inference procedure.

For ordinary principal components there is an intimate connection between principal components and factor analysis. This is somewhat less straightforward, using asymptotic arguments, in the dynamic case. It is explained in considerable detail in Section 3.1 of \cite{hall:lipp:2013}. That paper is one of a series of influential and much cited papers by Forni, Hallin, Lippi and Reichlein. The first paper, where fundamentals are explained, is \cite{forn:hall:lipp:reic:2000}. In that paper, so-called generalized dynamic factor models are introduced. We refer to the introduction of that paper for a historic account of developments leading up to generalized dynamic factor models, including papers by 
\cite{sarg:sims:1977}, \cite{gewe:1977}, \cite{cham:1983}, \cite{cham:roth:1983}, \cite{stoc:wats:2002}.  The model proposed in \cite{forn:hall:lipp:reic:2000}, and elaborated on in a number of follow-up papers referenced in \cite{hall:lipp:2013}, can be stated as
\begin{equation}
X_{jt} = b_{j1}(L)u_{1t}+b_{j2}(L)u_{2t}+\cdots b_{jm}(L)u_{mt}+\xi_{jt},
\label{2.4}
\end{equation}
where $\{X_{jt}, j =1,\ldots,p, t =1,\ldots,T\}$ is the original collection of time series, $b_{j1}(L),\ldots,$ $b_{jm}(L)$ are polynomials (or more general infinite MA components) in the lag operator $L$ associated with each of $m$ orthogonal white noise factor time series $\{u_{jt}\}$; the common components. The series $\{\xi_{jt}, j =1,\ldots,p, t =1,\ldots,T\}$ are series of so-called idiosyncratic components, which are zero-mean vector time series such that $\xi_{jt}$ is orthogonal to $u_{i,t-k}$ for any $i,j,t$ and $k$. The authors provide identification conditions, propose an estimator for the common components, and prove convergence results as both $p$ and $T$ tend to infinity. They use the model to construct a coincident index for GDP for countries in the European Union. Some very recent related papers are  \cite{chen:yang:zhan:2022} and \cite{yu:he:kong:zhan:2022}. 

A theoretical challenge with the dynamic factor models stated in \eqref{2.4} is that one has to let $p$ and/or $T$ tend to infinity in some of the theoretical derivations of the model. For example, these factor models are only asymptotically identifiable. If one is willing to simplify the models, at least some of these difficulties can be avoided. A fine example of such a modification is the paper \cite{lam:yao:2012}. They propose a model of form 
$$
X_t = {\bf A}Y_t+\varepsilon_t
$$
where {\bf A} is a $p \times m$ dimensional matrix, $\{Y_t\}$ is a $m$-dimensional latent factor process and $\{\varepsilon_t\}$ is a $p$-dimensional vector white noise process. They obtain asymptotic properties of estimators under two settings: (i), where $T \to \infty$ with $p$ fixed and (ii) when both $T$ and $p$ tend to infinity.

In spite of the success of dynamic factor type models, these models remain linear models with dependence based on second order moments such as variances and covariances. (A possible untried way out of this difficulty is to replace covariances and cross covariances with corresponding local covariances and cross covariances as in \cite{tjos:otne:stov:2022}.) In addition the dynamic factor models do depend on solving an eigenvalue problem which may turn out to be a prohibitive task if the dimension of $\{X_t\}$ is very large. Fortunately, there are alternative nonlinear embedding procedures that can be tried out as an alternative. These were surveyed in the static case in TJL. The dynamic case will be treated in the next section.

\section{Nonlinear dynamic embeddings \label{Nonlinear}}

\subsection{The methods \label{Method}}

We start by briefly listing the nonlinear embedding methods. Each method has been developed with a specific nonlinear situation in mind, and none of them work equally well in all situations. Presently we just mention the main principle underlying the construction of each method. Much more details are given in TJL. The dynamic extension for some of them are treated subsequently.

For the principal curve method, \cite{hast:1984}, the data are supposed to be concentrated roughly on a curve or more generally on a submanifold. Although the data in this case are not well represented by a linear model, they may still be well approximated by a local linear model giving rise to the LLE method of \cite{rowe:saul:2000} and to ISOMAP \citep{tene:silv:lang:2000}. For both of these methods, the distance between original data points is sought preserved in the embedding. Distance preservation is the dominating principle in the classical nonlinear method of multidimensional scaling -- MDS \citep{torg:1952}, and in fact ideas from MDS have served as a basis for several of the more recent nonlinear embedding methods. Not the least this is the case with random projections which originated from the famous \cite{john:lind:1984} lemma and which combines linear and distance preserving methods, and which avoids a data based embedding transformation.

\begin{figure}[ht!]
  \begin{center}
      \includegraphics[width=1.35\textwidth,angle=90,origin=c]{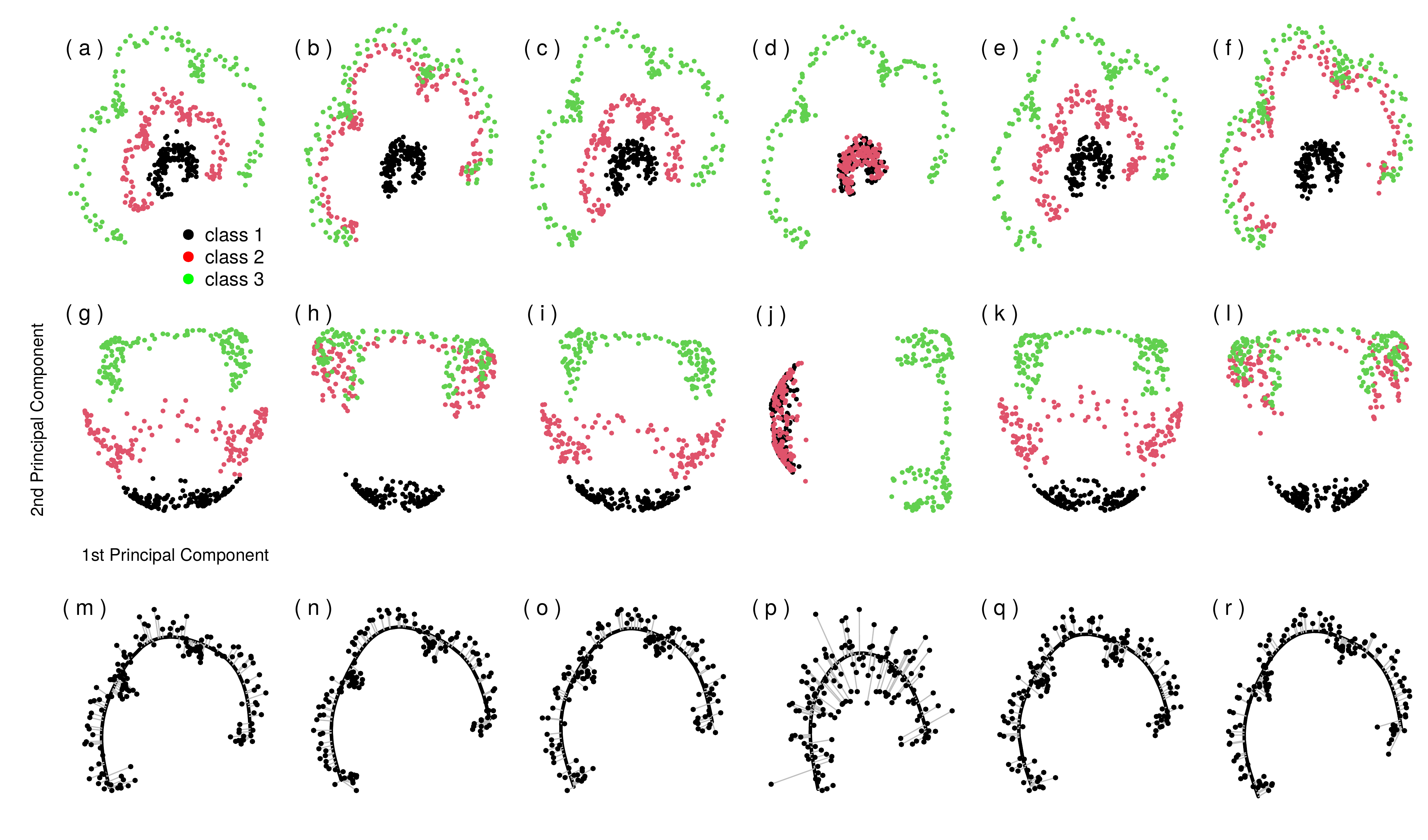}  
\end{center}
\caption{
  a: Three parametric curves from the so-called Ranunculoid, perturbed by Gaussian noise with a standard deviation of \sfrac{1}{2}.
  b-f: The middle Ranunculoid curve oscillates between the innermost and outermost curves, five time points are shown here.
  g-l: Kernel principal components (with a Bessel kernel) of a-f.
  m-r: Principal curves on the middle curves of a-f.
  \label{fig:1}}
\end{figure}

\afterpage{\clearpage}

A case which is particularly difficult to handle for PCA is the case where data lie on chained non-convex structures as for instance in Figure 1a. In this figure we present a data set that will be used for illustration purposes throughout and also in Section \ref{Dynamictopological} on topological data analysis. The raw data of Figure 1a consists of parts of three parametric curves, each being obtained from the so-called Ranunculoid, but with three different parameter sets. In addition, the curves have been perturbed by Gaussian noise. 
%Figure 1a corresponds to Figure 2a in TJL. 
For such and similar structures one may try to map the dependence properties to a graph or network (we use graph and network interchangeably in this paper), leading to a Laplace eigenvalue problem \citep{belk:niyo:2002} based on the adjacency matrix of the graph (this is a matrix constructed from the weighted distances between the data points constituting the graph). This approach is continued and extended  to diffusion maps \citep{coif:lafo:2006}. 

In still other situations it may be advantageous to use a nonlinear transformation of the data points prior to embedding, and then solve a resulting eigenvalue problem for the the embedded variables, as is done in kernel principal components in an associated reproducing kernel Hilbert space, see e.g.\ \cite{scho:smol:mull:2005}.

All of these methods are explained in more detail in TJL, and most of them illustrated by applying them to simulated data as in Figure 1. 
An additional method is Independent Component Analysis (ICA). The main concepts of the method are treated in a much cited paper by \cite{hyva:oja:2000}. It is related to PCA but with independent components instead of uncorrelated ones. Two other ones are autoencoding \citep{hint:sala:2006} and self organizing maps \citetext{\citealp{koho:1982}; \citealp[Chapter~14.4]{hast:tibs:frie:2019}}. The former will in fact play an important role in the present paper.

A very important asset of PCA is the relationship of PCA to factor analysis, where there exists a simple eigenvalue ratio-criterion for choosing the number of factors, or the number of principal components to keep in the general embedding. It should be noted that something similar to factor analysis is much harder to establish in nonlinear type embeddings. Eigenvalues can still be used to decide the dimension of the embedding for eigenvalue-based routines, but the interpretation is less straightforward. And for other nonlinear methods there seems to be no obvious way of determining the embedding dimension, meaning that it is often done by a trial and error routine.

%In TJL, and in general in embedding of iid data, a main application is clustering and classification. 
The most common application of embeddings of iid data is clustering and classification.
The embedding often results in a lower dimensional space, and clustering is more easily done in this space than in the original high-dimensional case. This is in particular the case if the embedding is done to a low-dimensional Euclidean space $\mathbb{R}^m$, to which standard clustering methods like k-means can be applied. 

For time series data as treated in the present paper, forecasting comes in as an additional and perhaps even more important problem. How does one forecast an ultra-high dimensional time series if one wants to include cross and auto dependence between the series. Lately artificial neural networks have played an increasingly important role in such a setting.
%Neural networks were also promoted in TJL as a speed saving device, but the so-called Skip-Gram method reviewed there used a simple layer neural network. Now, increasingly deep neural networks with particular structures are used especially in time series prediction. 
The commonly used Skip-Gram method is a only simple single-layer neural network. 
However, the use of deep neural networks with particular structures are increasing in usage also in this domain, especially for time series prediction.

To make the paper more self-contained a brief review of recent developments in artificial neural networks is contained in the Supplement \citep{suppJTSA}.

The neural network models, as seen in the Supplement, can contain a rather large number of unknown parameters to be learned during training. The possibility of overfitting is very real and quite often a regularization term is added to penalize too many parameters. This is in a way analogous to the use of a penalty term in the AIC criterion for time series. We give examples in Section \ref{forecasting}, equations \eqref{3.1} and \eqref{3.2}.

We are now ready to look at time-varying nonlinear embedding in general and at forecasting in particular.

\subsection{Time-varying nonlinear embedding \label{Time-varying}}

In the beginning of this section we listed a number of embeddings methods.
% many of which we analyzed in TJL including an illustrating example.
How do these methods hold up in the presence of dynamics in form of time variations? We have to distinguish between two types of time variations: stationary or nonstationary. In the stationary situation, if observed for a long enough time interval, one expects that one  will have a invariant embedding pattern as time goes on. 

The algorithms for the methods of MDS, ISOMAP, LLE, spectral graph methods, diffusion maps, random projections and kernel principal components all work if time variations are neglected in  the sense that computations are done separately for each time point. They may produce useful results but the foundation that these results build on, is uncertain. This has been illustrated by \cite{zhan:tong:2021}  in the case of asymptotic analysis of eigenvalues and eigenvectors if ordinary principal components are applied in a time series situation. We are not aware that any corresponding analysis has been undertaken for the nonlinear embedding methods mentioned in this paragraph.

For principal components there exists a modification of the method in the so-called dynamic principal components and dynamic factors initially based on the work by \cite{bril:1975} in the spectral domain. This was reviewed in Section \ref{Principal}, and has been used in many successful applications. Again, to the best of our knowledge, there is no work in this direction for the nonlinear embeddings mentioned in the previous paragraph. Certainly, if analogue approaches could be found it would be of considerable interest. A very different alternative may be to take the Brillinger spectral approach in another direction, by using either a higher order spectrum or a local spectrum, cf.\ \cite{jord:tjos:2022}, as a point of departure for a nonlinear dynamic principal component analysis.

As far as embedding in a time-varying framework is concerned, it may be advantageous to use a neural network based learning approach. The reason is that in the training process the time sequence of the observations are taken into account. (This dependence on time may even be nonstationary.) A prime example of embedding in using neural networks is the much cited paper by \cite{hint:sala:2006}, where they use autoencoding as a tool. 

Autoencoding in its simple basic form has two main parts: an encoder that maps the input into the code, and a decoder that maps the code to a reconstruction of the signal. Essentially it is a one hidden layer feedforward neural network,  where the output y-variable in principle should be identical to the input x-variable, but where in practice autoencoders are typically forced to reconstruct the input approximately, preserving only the most relevant aspects of the data. The low dimensional embedding is constituted by the hidden layer representation. The autoencoders have been further developed from this simple one hidden layer structure. Some of the most powerful and recent artificial intelligence methods involve autoencoders stacked inside deep neural networks, see e.g.\ \cite{domi:2015}.

Autoencoding is a data analytic or machine learning technique that has turned out to be instrumental in dimension reduction. But unlike dynamic principal components it does not rest on a traditional mathematical statistics foundation. It contains parameters, the embedding dimension being one of them, that must be chosen, and it may not be obvious how an optimal choice can be made. This potential lack of a statistical model-based routine to estimate parameters in an algorithm accentuates a difference between a machine learning approach and a mathematical statistics approach. We refer to TJL for a further discussion of this.% where the differences are summed up in three keypoints.

When it is clear that the time variations are nonstationary, one has to take this into account in the embedding. Then, unlike the stationary case, the embedding can be expected to change significantly with time. There are some papers on this, largely empirical in nature. Time-varying multidimensional scaling is examined in \cite{lope:mach:2014}, \cite{he:shan:xion:2018}. Time-varying principal curves are looked at by \cite{li:gued:2021}, whereas two types of ISOMAP streaming are considered in  \cite{maha:chan:2020}. \cite{lian:talm:zave:cari:coif:2015} study diffusion maps using a Kullback-Leibler criterion. Another application of diffusion maps to periodic physiological data is in \cite{lin:mali:wu:2021}.

To our knowledge there is no systematic comparison of the embedding methods under time varying circumstances.  Can one find classes of nonstationarity and corresponding classes of ``optimal'' embedding methods? Are some embedding methods more robust than others to say periodic disturbances or to outliers? This seems to be a wide open research field. 

In TJL we illustrated some main embedding algorithms on a data set consisting of chained Ranunculoid curves with added Gaussian noise, as depicted in Figure 1a. Here we develop this illustration further by examining how the various methods react to changes in the data structure. We let the Ranunculoid curves go through a sinusoidal movement. More precisely, we let the middle Ranunculoid curve oscillate between the curves 1 and 3 as depicted in Figure 1b-1f, where 5 time points are displayed. In addition, different seeds have been used in the generation of the point cloud for each time point. The present example could be considered as a very simple toy example motivated by physiological experiments that quite often has an oscillatory structure. 

In the static case of TJL we concluded that PCA does not work for such a chain of non-convex curves, and it is therefore meaningless to try to use PCA to describe the temporal data in Figure 1b-1f. However, the nonlinear embedding methods mentioned in the current section have the potential to pick up this oscillatory movement. This is illustrated by the kernel principal component method in Figure 1g-1l. It is seen that it gives a faithful embedding representation of the periodic pattern of the Ranunculoid structures. The embedding is rotated 90 degrees in Figure 1j, but it is unclear why the algorithm does that. As can be expected the kernel principal component method is not able to distinguish between all three curves when the middle Ranunculoid curve is very close to the innermost or the outermost curve.

In Figure 1m-1r we have also displayed the time variation for the principal curve estimated from the moving middle curve in the Ranunculoid. It is seen that the shape of the principal curve is invariant, but as expected the scale varies, and the noise is relatively largest when the middle curve is most compressed.

The choice of the kernel principal component method in Figure 1 is completely arbitrary. Similar results are obtainable for the LLE and ISOMAP methods used in TJL in the static case. Those methods are also able to pick up the temporal changes. As indicated earlier in this section, there is not much literature on nonlinear time varying embeddings, and there seems to be room for more analytic and empirical work.

\begin{figure}[ht!]
  \begin{center}
      \includegraphics[width=0.58\textwidth]{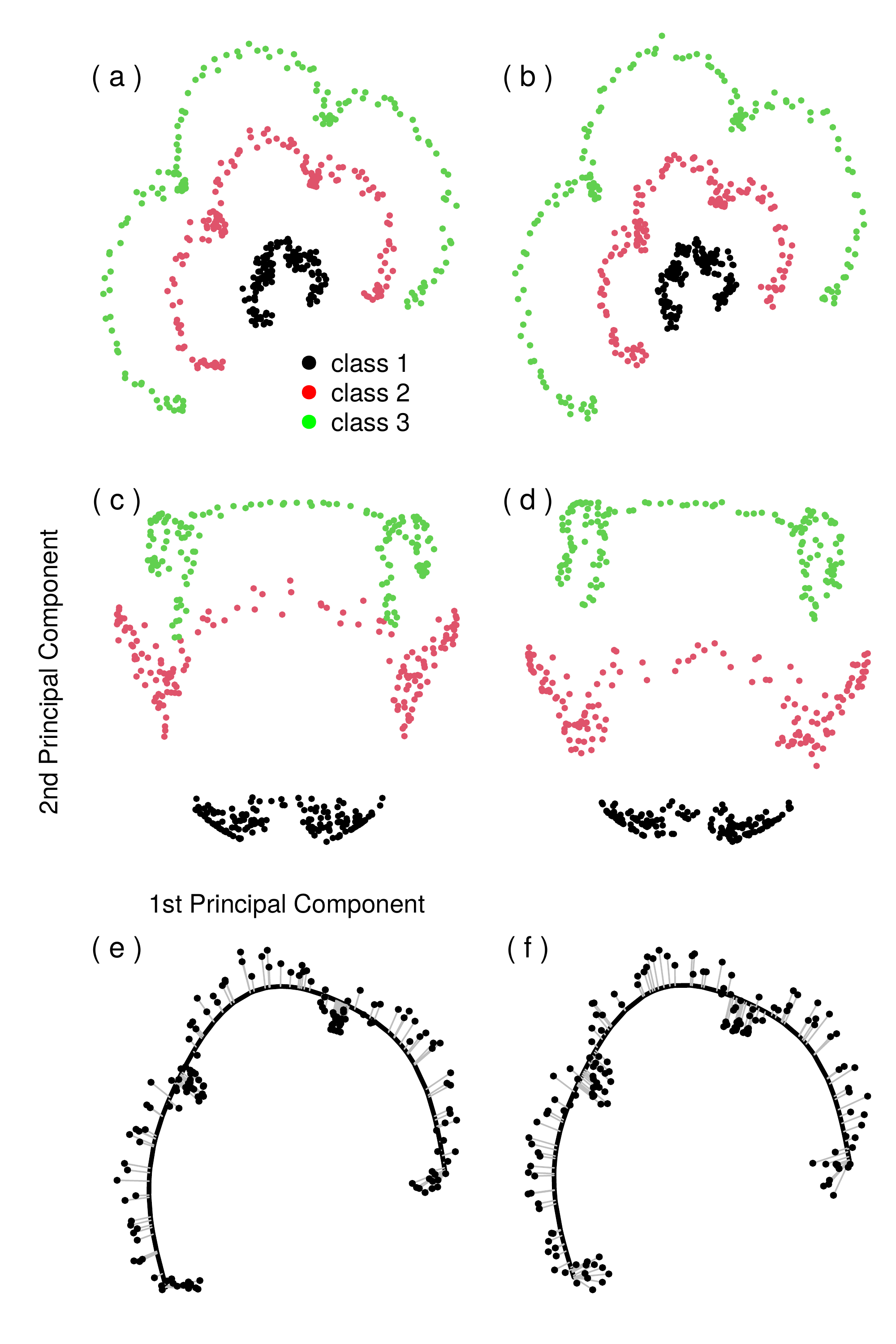}  
\end{center}
\caption{
  a and b: Average of the first three (a-c) and last three (d-f) time points from Figure \ref{fig:1}.
  c and d: Kernel principal components (with a Bessel kernel) of a and b.
  e and f: Principal curves on the middle curves of a and b.  
  \label{fig:2}}
\end{figure}

Sometimes it is recommended to take averages to recover the main structure of a sequence of embedding plots. This has been done in Figure 2a-2d, where Figure 2c represent the embedding of the average of the point clouds in Figures 1b-1d, and Figure 2d represents the average of Figure 1d-1f. It is seen that in this very simple case the principal embedding pattern is recovered, especially in the latter case, chiefly due to the better separation of the innermost curves in Figure 1e as compared to Figure 1c.

\subsection{Forecasting and embedding \label{forecasting}}

In TJL, the time dimension was neglected, and the embedding was primarily motivated by clustering and classification. In time-varying or dynamic problems an added motivation for embedding is forecasting.

In finance, but also in other areas, there is a need to forecast very high dimensional time series. For example, it may be necessary to obtain simultaneous forecasts of a very large collection of time series. As such time series are often cross and autocorrelated, one ideally wants to take advantage of the joint history to produce better forecasts. However, use of traditional and classical forecasting methods such as vector autoregressive, GARCH, exponential smoothing and state space modeling \citep{hynd:koeh:ord:snyd:2008,hynd:atha:2018} may not be expected to work well when the dimension radically increases.

One alternative to avoid this problem is simply to make individual univariate predictions for each time series. This means losing all cross dependence information in the joint history of the process. 

A compromise is to use embedding of the time series to a manageable dimension, then make a forecast using for instance one of the standard methods for the embedded series and transform back again.
This is straightforward if a principal component (or even principal dynamic factor) is used, and the dimension is not so high that the eigenvalue problem creates trouble. Indeed, an advantage of using principal components is that it is easy to transform back to the original time series, since everything is linear. For the nonlinear case the back-transformation problem is far more serious. This has been examined by e.g.\ \cite{papa:talm:kevr:siet:2021}. 
They use locally linear embedding (LLE) and diffusion maps as their two embeddings routines, and with radial basis function interpolation and geometric harmonics to create back-transformations.

There are also a number of papers where neural networks methods are used. First, let us consider the linear  temporal model by \cite{yu:rao:dhil:2016}.
One reason for selecting this paper is its transparent handling of the regularization mechanism to avoid overfitting of the embedding model. We will use the notation in \cite{nguy:quan:2021} in describing this. 

Consider a multivariate time series $X_t, \ t =1,\dots,T$ of dimension $p$, where $p$ may be in the range of $10^3$ to $10^6$, or even higher. The collection of time series  forms a $T \times p$ dimensional matrix ${\bf X}$. By using an embedding, this matrix is changed to a $T \times m$ dimensional matrix ${\bf Y}$, where $m$ is the embedding dimension. To train a neural network like RNN (Recursive Neural Network, see the Supplement for a definition), data can be batched temporarily. Denote by ${\bf X}_B$ a batch of data containing a subset of $b$ samples ${\bf X}_B = \{X_t,\ldots,X_{t+b-1}\}$, where $B =\{t,\ldots,t+b-1\}$ are time indices. \cite{yu:rao:dhil:2016}, as described by \cite{nguy:quan:2021}, perform a constrained linear embedding or regularization by solving
\begin{equation}
\min_{{\bf Y},{\bf F},{\bf W}}{\cal L}({\bf Y},{\bf F},{\bf W}) = \frac {1}{|{\cal B}|}
\sum_{B \in {\cal B}} {\cal L}_B({\bf Y}_B,{\bf F},{\bf W}),
\label{3.1}
\end{equation}
where ${\cal B}$ is the set of all data batches and each batch loss is defined by
\begin{equation}
{\cal L}_B({\bf Y},{\bf F},{\bf W}) = \frac{1}{nb}||{\bf X}_B-{\bf F}{\bf Y}_B||_{l_{2}}^2+\lambda {\cal R}({\bf Y}_B,{\bf W}).
\label{3.2}
\end{equation}
Here, ${\cal R}({\bf Y}_B,{\bf W})$ is a regularization of ${\bf Y}_B$ parameterized by ${\bf W}$ to enforce certain regular properties of the latent terms, and $\lambda$ is a regularization parameter. The regularity that \cite{yu:rao:dhil:2016} impose is to assume that $\{Y_t\}$ follows a vector autoregressive model so that $Y_t = \sum_{j=1}^ {L}{\bf W}^ {(j)}Y_{t-j}$ where $L$ is a predefined lag parameter. Then the regularization becomes
$$
{\cal R}({\bf Y}_B,{\bf W}) \doteq \sum_{l = L+1}^{b} || Y_l - \sum_{j=1}^l{\bf W}^{(j)}Y_{l-j}||_{l_{2}}^2.
$$
The optimization is then solved via alternating minimization with respect to the variables ${\bf Y}, {\bf F}$, and ${\bf W}$.

This example illustrates the regularization for a linear model, but essentially the same method can be used for a nonlinear embedding and for a neural network type regularization. \cite{nguy:quan:2021} use autoencoding as an embedding device. The embedded time series process of lower dimension is subsequently forecasted using a neural network LSTM forecaster as outlined in the Supplement. 

Again  parameters such as the embedding dimension, the lag structure, the regularization parameter and the parameters entering in the LSTM algorithm have to be chosen or estimated. It does not seem to be clear how an optimality theory for this procedure can be established or how sensitive the embedding and the resulting forecast are to these choices.

Until recently, traditional forecasting methods such as ARMA-based and exponential smoothing \citep{mcke:1984}, and state-based models have consistently outperformed machine learning methods such as RNNs in large scale forecasting competitions \citep{makr:hynd:petr:2020,makr:spil:assi:2020,cron:hibo:niko:2011}. A key reason for recent successes of deep learning in forecasting is multitask univariate forecasting, sharing deep learning parameters across all series, possibly with some series-specific scaling factors or parametric model components \citep{sali:flun:gast:2019,smyl:2020,band:berg:hewa:2020,wen:tork:nara:made:2017}.
Indeed, the winner of the M4 forecasting competition of \cite{makr:spil:assi:2020}, involving 100 000 time series and 61 forecasting methods, was a hybrid exponential smoothing-RNN model \citep{smyl:2020} in which a single shared univariate RNN model is used to forecast each series, but seasonal and level exponential smoothing parameters are simultaneously learned to normalize the series. This is an example of a very successful interaction of machine learning methods with a traditional parametric forecasting modeling. Clearly, this is an inspiration for more of this interaction in other problems like clustering and, later, embedding of networks.  

It should also be noted that much work in forecasting has been concentrated on point forecasting. As argued by among others \cite{nguy:quan:2021}, low-dimensional embedding is probably an important tool to obtain a probability distribution of forecasts. More work is required in this area!

\section{Dynamic topological data analysis \label{Dynamictopological}}

The field of topological data analysis (TDA) is new. It has emerged from research in applied topology and computational geometry initiated in the first decade of this century. Pioneering works are \citet{edel:letc:zomo:2002} and \citet{zomo:carl:2005}. \citet{chazal2021introduction} give a relatively nontechnical review which is also oriented towards statistics, and a short summary can be found in Section 4.2 of TJL. We include a few main points from that paper in the following.

For our purposes of statistical embedding, TDA brings in some new aspects in that topological properties are emphasized and can potentially be used as new characterizations of the data cloud. An important device is the so-called persistence diagram which depicts the persistence, or lack thereof, of certain topological features as the scale in describing data cloud changes.

Assume that $n$ data points ${X_1,\ldots,X_n}$ are at or close to a smooth compact submanifold $S$. One may estimate $S$ by trying to cover the data cloud by a collection of balls of radius $\varepsilon$, such that 
\begin{equation}
\hat{S} = \cup_{i=1}^n B(X_i,\varepsilon),
\label{4.1}
\end{equation}
where $B(X_i,\varepsilon) = \{x: ||x-X_i|| \leq \varepsilon\}$. One may question what happens to this set as the radius of the balls increases. Consider for example a data cloud that contains a number  $n$ of isolated points that resembles a circular structure. Let each point be surrounded by a neighborhood consisting of a ball centered at each data point and having radius $\varepsilon$. 
Then, initially and for a small enough radius $\varepsilon$, the set $\cup_{i=1}^{n} B(X_i,\varepsilon)$ will consist of $n$ distinct connected sets (of so-called homology zero). But as the radius of the balls increases, some of the balls will have non-zero intersection, and the number of connected sets will decrease. For $\varepsilon$ big enough one can easily imagine that the set $\cup_{i=1}^{n} B(X_i,\varepsilon)$ is large enough so that it covers the entire circular structure obtaining an annulus-like structure of homology 1, but such that there still may exist isolated connected sets (homology 0) apart from the annulus. Continuing to increase the radius, one will eventually end up with one connected set of zero homology. 

This process, then, involves a series of births (at $\varepsilon$-radius zero $n$ sets are born) and deaths of sets as the isolated sets coalesce. 
A  useful plot is the  persistence diagram, which has the time (radius) of birth on the horizontal axis and the time (radius) of death on the vertical axis. 
The birth and death of each feature is represented by a point in the diagram. All points will be above, or on the diagonal, then. For the circle example mentioned above, the birth and death of the hole will be well above the diagonal, and it has a time of death which may be considerably larger than its time of birth. The birth and death points of the connected components, on the other hand, may be quite close to the diagonal if the distances between points are small enough. 

In TJL we have gone through this process for the data set of the three chained Ranunculoids displayed in Figure 1a of the present paper, and the reader is referred to Figures 3 and 4 of that paper for a description of the persistence diagram for two levels of noise. 

The idea is that this description of a  point cloud in the plane, as
indicated above,  may be generalized to higher dimensions and much
more complicated structures with multiple holes and voids of
increasing  homology. The number of sets of different homologies are
described by the so-called Betti numbers, $\beta_0, \beta_1,
\ldots$. In a non-technical jargon $\beta_0$ is the number of
connected components ($\beta_0 = n$, $n$ being the number of isolated
points in the start of our example), $\beta_1$  is the number of
one-dimensional holes, so $\beta_1= 1$ if there is only one connected
ring structure, and $\beta_0 = 1, \beta_1=0$ when the radius is so
great that there is only one connected set altogether. The hole is
one-dimensional since it suffices with a one-dimensional curve to
enclose it, whereas the inside of a soccer ball is two-dimensional; it
can be surrounded by a two-dimensional surface, and has $\beta_0 =1,
\beta_1=0$ and $\beta_2=1$. A torus has $\beta_0=1, \beta_1=2,
\beta_2=1$.

\subsection{Forms of dynamics in TDA \label{Forms of dynamics}}

The simplest form of investigating the dynamics is just to analyze a collection of high dimensional time series, say, by taking snapshots at several time points and find the persistence diagram for the data cloud of each snapshot. In this series of persistence diagrams one may look for key features like type of periodicities that may be difficult to obtain by other means. As mentioned, periodicities is a main feature of several physiological processes, for instance in the brain, see e.g.\ \cite{chun:huan:caro:calh:gold:2022}. 

\begin{figure}
	\begin{center}
		\includegraphics[width=1.30\textwidth,angle=90,origin=c]{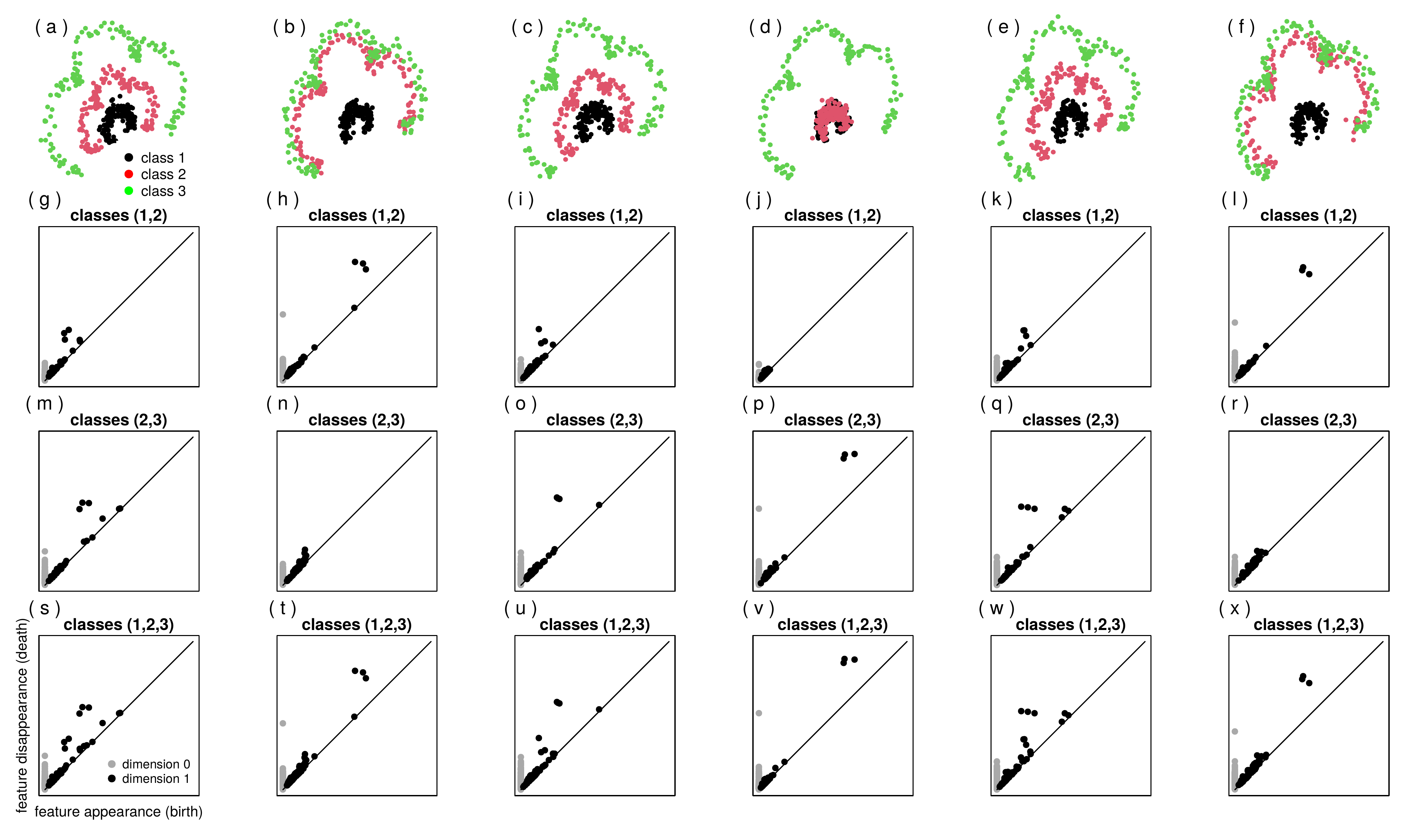}  
	\end{center}
	\caption{a-f:  These correspond to the plots in Figure \ref{fig:1}.
		g-x: Persistence diagrams for classes (1,2), (2,3) and (1,2,3) for a-f.
		\label{fig:3}}
\end{figure}

Time changes in the persistence diagram can be illustrated by the previous Ranunculoid example. This has been depicted in Figure \ref{fig:3} for the same periodic time variation of the Ranunculoid curves as in Figure 1. The Ranunculoid plots of Figure 3a-3f correspond to the plots in Figure 1a-1f. Here, Figure 1a, the static case, corresponds to Figure 4a in TJL (using different seeds though). The persistence diagrams in the present paper for the static case in Figure 3g for classes (1,2) and in Figure 3m for classes (2,3) and in Figure 3s for classes (1,2,3) correspond to Figures 4e, 4f and 4h of TJL. 

The gray points represent sets of homology zero (isolated sets), and the black points represent sets of homology one; i.e. one dimensional holes. The gray columns at the left is just the time of death for all the sets around the individual points as the radius $\varepsilon$ of expression \eqref{4.1} for the individual points increases. Naturally, for the static picture the gray column for classes (2,3) reaches higher than for classes (1,2) since the distance between classes 2 and 3 is larger. The black points at the right of the gray column represent small holes that form as $\varepsilon$ increases. They are due to indents in the point spreads of the Ranunculoids. If the Ranunculoids were to be replaced by noiseless circles having the same center they would disappear. By comparing the diagrams in Figure 3g and Figure 3m, it is seen that the holes have larger lifetimes for the (2,3) class, which is natural in view of the larger indents in curve 3 in Figure 3a.

The temporal patterns of the persistence diagrams that appear as time progresses are quite intuitive. For example, corresponding to the proximity of curves 2 and 3, and the larger distance between curves 1 and 2 in Figure 3b and Figure 3f, there are few holes of any length of lifetime in diagrams 3n and 3r, whereas there are three holes of sizable lifetime in Figures 3h and 3l, corresponding to the three main indents of curve 2 for the (1,2) diagram. As the radii $\varepsilon$ increase, the two curves coalesce, and we have a death at the gray point above the gray column in the (1,2) diagram in Figures 3h and 3l. 

For the case of Figure 3d where the curves 1 and 2 are very close, the situation is reversed as can be seen from Figures 3j and 3p. The other intermediate situations in Figures 3c and 3e give intermediate persistence diagrams. Finally, the case of (1,2,3) taken simultaneously: Because of the relative simple situation of Figure 3, the persistence diagrams are virtually superpositions of the diagrams for (1,2) and (2,3).

The point of all this is that the time variation of the persistence diagrams reveals temporary topological variation which is not apparent from the nonlinear plots associated with the methods of Section 3, as it is partially revealed in Figures 1 and 2.

To compare features originated from specific persistence diagrams, a distance measure between such diagrams is needed. Several such distance measures exist; perhaps the most well-known is the bottleneck distance. Given two diagrams $C_1$ and $C_2$, the bottleneck distance is defined by
$$
\delta_{\infty}(C_1,C_2) = \inf_{\gamma}\sup_{z \in C_1}||z-\gamma(z)||_{\infty},
$$
where $\gamma$ ranges over all bijections between $C_1$ and $C_2$. Intuitively, this is like overlaying the two diagrams and asking how much one has to shift the diagrams to make them the same \citep{wass:2018}. The practical computation of the bottleneck distance amounts to the computation of perfect matching in a bipartite graph for which classical algorithms can be used \citep{chaz:mich:2017}.

An alternative to the persistence diagram is to use the so-called persistent landscapes \cite{bube:2015} which has the advantage that it is a function space. The bottleneck distance is also a natural tool in statistical inference on persistent landscapes, cf.\ \cite{chaz:fasy:lecc:rina:wass:2015}.

The snapshot procedure is a primitive procedure, that although it can illustrate nonstationary variations of a point cloud, it does not really attempt to model the individual time series whose simultaneous observations create the point cloud. A quite different approach that tries to take individual time series structure into account is using the Takens' embedding procedure of a time series. 

Consider a (one-dimensional) time series $\{X_t,t=1,\ldots,T\}$. Takens' embeddings procedure can be used to convert the time series into a point cloud with points $v_i = \{X_i,X_{i+\tau},\ldots,X_{i+(d-1)\tau}\}$, where $d$ specifies the dimension of the embedding frame of the points and $\tau$ a delay parameter.
Taking $d=2$ and $\tau=1$ yields a point cloud in the plane with $v_i = \{X_i,X_{i+1}\}$. This scatterdiagram in the plane is often used to illustrate the dynamic properties of a time series, and has in particular been used to illustrate limit cycles for nonlinear time series \citep{tong:1990}. In general both $d$ and $\tau$ will have to be determined in practice. The resulting persistence diagram would furnish a topological characterization of the time series. Again it may be of interest to seek for topological type periodicities by means of the persistence diagram. One may be able to detect periodicities not revealed by the power spectrum, see also \cite{jord:tjos:2022}. It is also possible to take the power spectrum of a time series as point of departure, and find the persistence diagram of it, by taking advantage of the algorithm for constructing a persistence diagram for a function \citep{ravi:chen:2019}.

It is not straightforward to generalize this approach to multivariate time series, but see ideas can be found in \cite{gide:katz:2018} and \cite{bour:chun:omba:2022}. The latter has its main focus on brain wave data collected over different channels.

In topological analysis of brain data, the data of the various channels are usually modeled as a brain network. For dynamically changing brain networks it is assumed that the node sets (channels) are fixed, while weights on links may be changing in time. If one builds persistent homology at each fixed time, the resulting persistence diagrams are also time dependent.

Actually, however, in more complicated situations, the persistence diagram is not computed directly from scale shifts as in the expression \eqref{4.1} but from so-called simplical complexes. This approach is particularly interesting since it generalizes the embedding of a point cloud in a graph as will be treated in the dynamic case in the next section. We give a brief description of simplical complexes in the Supplement. Much more details can be found in  \cite{chaz:mich:2021}.

There are a number of problems of interest for statisticians in general and for time series analysts in particular in TDA. \cite{chaz:mich:2021} mention four topics in the static case, but with self-evident interest in the dynamic TDA case:
\begin{enumerate}
\item Proving consistency and studying the convergence rates of TDA methods.
\item Providing confidence regions for topological features and discussing the significance of estimated topological quantities.
\item Selecting relevant scales (i.e.\ selecting $\varepsilon$ in the examples discussed above and in the Supplement) at which topological phenomena should be considered as functions of observed data.
\item Dealing with outliers and providing robust methods for TDA.
\end{enumerate}
\cite{chaz:mich:2021} have used the bootstrap in a static TDA situation. One may want to introduce the block bootstrap to take better care of dependence structures. There are also recent contributions to hypothesis testing, \cite{moon:laza:2020}, sufficient statistics, \cite{curr:mukh:turn:2018}, and Bayesian statistics for topological data analysis, \cite{maro:nasr:obal:2020}.

It has been seen that construction of neighborhood graphs and generalization of these are important tools in TDA and elsewhere. In the next section we look at the situation where the data are given in terms of a graph and where time variations are included.

\section{Dynamic graphs \label{Dynamic graphs}}

In the preceding sections we have seen how graphs can be useful tools in nonlinear embeddings of a point cloud, and in TDA in handling of a point cloud when using simplical complexes (cf.\ the Supplement for more details). In the present section it is assumed that the data themselves are given in the form of a network. With the increasing use of the internet and Big Data, analysis of large networks is becoming more and more important There is a wide field of applications ranging over such diverse areas as e.g.\ finance, medicine and sociology, including criminal networks. A broad overview can be found in the recent book by \cite{newm:2020}. More foundational problems are covered in \cite{cran:2018}.

\subsection{The static case \label{Static case}}

Both research and applications have been overwhelmingly concentrated on static networks. But a change is presently taking place, since the static assumption in many cases is an untenable one. In many types of networks, as time goes on, new nodes are added to the network, others are vanishing, and the strength of the connection between nodes are changing, or may even be severed. This realization has led to a rapid recent increase in the interest for dynamic networks. In this paper we will only be able to give a glimpse of this development, but it harbors several open and exciting problems for time series analysts. To put this into context, however, we first need to give a very brief overview of methods for static networks. This is from TJL and \cite{kaze:goel:jain:koby:seth:fors:poup:2020}.

%The purpose of TJL was to review embedding theory generally and for networks in particular. 
By embedding a network in Euclidean space $\mathbb{R}^m$ or on a submanifold in $\mathbb{R}^m$, the nodes of the network are represented by vectors on which further analysis like clustering and classification can be undertaken. 
%In TJL, we essentially looked at two methods, 
From our point of view, two methods stand out as particularly interesting: Spectral embedding and embedding via the so-called Skip-Gram neural network method.

Spectral embedding is motivated by the clustering problem where clusters form communities in the network. The problem is to find these communities. This is done by minimizing a cut functional or maximizing a modularity functional. In both cases the minimizing/maximizing leads to an eigenvalue problem analogous to the Laplace eigenvalue problem  mentioned in Section \ref{Method}. A faster solution of the modularity problem is obtained by the so-called Louvain method for community detection.

It is computationally costly to solve an eigenvalue problem for high dimensions, and networks often have ultra-high dimensions. These problems are to a large degree alleviated in the neural network-based Skip-Gram procedure. The Skip-Gram procedure was first developed in word embedding in a language text (from this the nomenclature ``Skip-Gram''). Here the eigenvalue problem is removed altogether, and the neural net training is speeded up by so-called negative sampling. The neural net in Skip-Gram has only one hidden layer. The input to the hidden layer is formed by linear combinations of the nodes, the output by linear combinations of the hidden units plus possibly a nonlinear sigmoidal type transformation of these output linear combinations to obtain probabilities. The idea is for each node to associate neighboring fragments of the network by performing random walks governed by the weights on the links of the neighboring nodes for the given node. These fragments are fed through the hidden units and the optimal linear combinations are learned by requiring essential conservation of the fragments as they are passed through the hidden units. The dimension of the hidden layer in the network is much lower (perhaps in the range from 500-600 or so). The output linear combinations are taken as representations (or embeddings) of the nodes. In principle all of the hidden layers has to be updated for each iteration with a new neighboring fragment. In practice this is avoided by the idea of negative-sampling where just a random selection of the hidden units are updated for each iteration. 

It has been found that the method works rather well in practice. The basic publications \citep{miko:chen:corr:dean:2013a,miko:suts:chen:corr:dean:2013b} have well over 20 000 citations. There are several versions of the method, and as will be seen, there is also a dynamic one, to be mentioned shortly. For more details and references the reader is referred to TJL.

The neural network approach has a natural extension to networks with several layers. We refer to the Supplement for a brief description of these, including autoencoding, convolution networks and recurrent networks, mainly from a prediction point of view, but they can equally be used in a static framework for embedding, clustering and classification.

The methods presented in this section can all be characterized as machine learning methods. There is really no statistical model involved. Nevertheless these procedures contain parameters or hyper parameters that have to be determined. The performance of the methods may depend quite critically on the choice of these parameters \citep{peix:2021}. They can be determined by optimizing an object type function in some cases, but in other situations one has resorted to more trial and error procedures. 

In an attempt to counter some of these problems, a more traditional statistical model has been introduced for community finding in networks. This is the stochastic block model (SBM). In the simplest undirected stochastic block model each of the nodes is assigned to one of $k$ blocks (communities) and undirected edges are placed independently between  node pairs with probabilities that are a function only of the block membership of the nodes. This results in $k^2$ probability parameters for the model, and there are several ways of estimating them for a given real data network. There are a number of papers in which asymptotic distributional properties and consistency are developed (see e.g.\ \cite{zhan:chen:2020}
and references therein). Unfortunately, however, the simple SBM does not work well for many real world networks. One relatively simple generalization is the degree corrected stochastic block model (dcSBM). It allows for heterogeneity in the number of degrees (essentially the sum of the weights on the links associated with a node) for the nodes. This is a phenomenon that is often observed in practice, and it seems to work much better on real life networks.

%The degree corrected \todo{Riktig?} SBM was illustrated on static networks in TJL.
%The embedding dimension was chosen to be $m=2$; i.e., visualization, for which three methods were presented.

\subsection{The dynamic case \label{Dynamic case}}

If we have a sequence of networks in time $\{(G_t,V_t),t=1,\ldots,T\}$, where $G_t$ denotes the nodes at time $t$, and $V_t$ the links at this time, the simplest procedure to analyze dynamics is to obtain a snapshot embedding for each time point $t$. If we just need one embedding for the entire time period, this can be simply obtained by taking averages of the adjacency matrices and finding the embedding corresponding to this average. Alternatively, but much more time consuming, one may find an embedding separately for each time point $t$, and then take the average of the embedding vectors (cf. Figure 2). If one wishes to give more weight to the most recent observations, this can be obtained by taking a weighted average. %\todo{Refer to figure}

However, there are a number of problems with this approach in addition to the fact that it may be close to impossible to carry out because of the time needed to produce each snapshot for large networks. It turns out that the embeddings we have mentioned so far, perhaps especially the Skip-Gram procedure, may be rather unstable in the sense that a relatively small change in the network from one time point to another may produce large changes in the embedded vectors, even for vector representation for nodes lying at a considerable distance from where the main changes are taking place, see the illustrations in Section \ref{sec:dyn_illustation}. This is illustrated in Figure \ref{fig:dyn_example_vis}. Regularization analogously to the formulas \eqref{3.1} and \eqref{3.2} has been suggested, but where the added penalty term penalizes big changes in time for the embedding vectors for the nodes.

For the spectral embedding approach it has been proposed that small perturbations could be handled by Taylor expansion. If the Laplacian matrix and the degree matrix are changed by small amounts, changes in $\Delta {\bf L}$ and $\Delta {\bf D}$ can be computed, from which updated eigenvalues and eigenvectors can be obtained. The Davis-Kahan theorem (cf.\ \cite{yu:wang:samw:2014}) gives an approximation error for the top $m$ eigenpairs. An alternative approach in the spectral representation is to stack the adjacency matrices ${\bf A}_1,\ldots,A_{T}$ into an order 3 tensor \citep{dunl:kold:acar:2010}. Such embeddings can be used to make predictions of links.

For the case of Skip-Gram, more precisely for the so-called LINE-version, \cite{du:wang:song:lu:wang:2018} propose a decomposition procedure, where new nodes can be taken into the objective function in a separate part. For each time step this can be used to obtain a vector embedding of the new nodes as well as those nodes that are most affected by the new nodes. There is a criterion for which nodes need a new representation. The authors claim a large time saving compared to a full analysis at each time point.

Relationships with time series networks generated by topological data analysis is explored in \cite{varl:spor:2022}. 
All of the above have concentrated on discrete time. There are also some papers where the dynamics of the network is expressed in terms of differential equations. See e.g.\ \cite{raj:cai:xie:pala:owen:mukh:naga:2020}.

All of these approaches are algorithmic in nature, extensions of algorithmic and machine learning algorithms used in the static case. How about the SBM model that was heralded as a contrast to the algorithmic methods in the static case? Is there a dynamic theory for it? Compared to the static case it seems to be quite limited. \cite{xu:hero:2014} have used the extended Kalman filter to describe the dynamics. \cite{ludk:eckl:neal:2018}  have an interesting approach where edges are switched on or off according to a hidden Markov chain.

\subsection{Dynamic SBM illustration}
\label{sec:dyn_illustation}

In TJL we illustrated the static SBM model on a simulated system containing two communities using various configurations. 
We also re-mapped high dimensional embeddings into 2-dimensional visualization space to visually look at the embeddings.
 In particular, we tried three visualization methods, the t-SNE, the UMAP and LargeVis methods being based respectively on statistical nonlinear embedding, TDA and Skip-Gram, and also compared to principal components. It is of interest, we think, to extend this illustration to a dynamic case. 

Ideally one would like a dynamic model where nodes could be added or removed in time or where the strength of edges could vary. To our knowledge such models have not really been analyzed in the SBM-literature, but such networks are clearly very realistic. As a rather simple minded and preliminary approach we have opted for a graph constructed by combining three SBM models, and where nodes are subsequently removed randomly one at a time. We think that this model, despite its simplicity, will bring forward some main problems inherent in dynamic graph modeling.

\begin{figure}[ht!]
	\begin{center}
		\includegraphics[width=\textwidth]{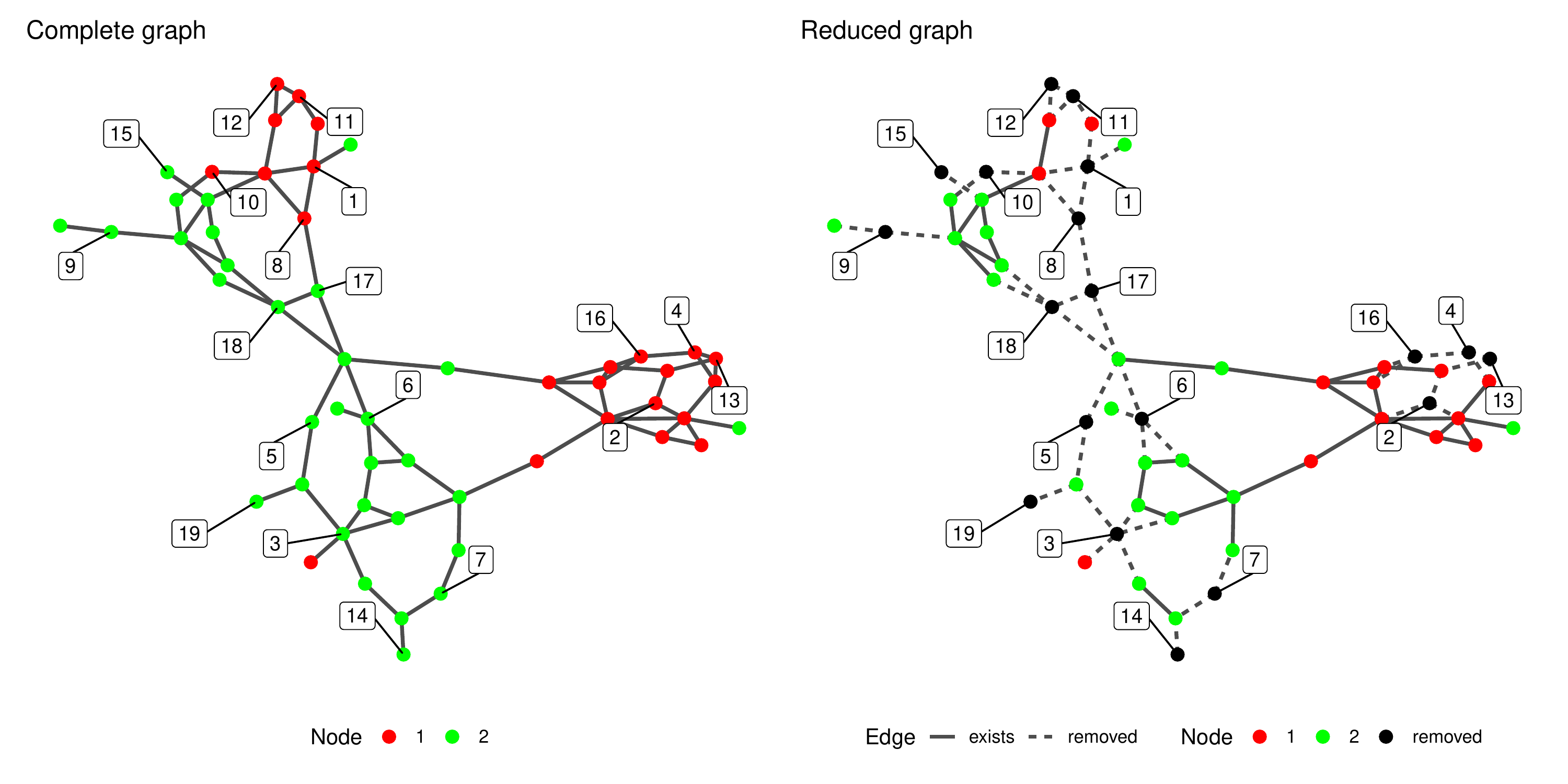}  
	\end{center}
	\caption{Illustration of graph used in the dynamic embedding illustration. 
		The left panel shows the complete graph before reduction. 
		The right panel shows the  graph in the 20th stage, after removal of 19 nodes (and their associated edges).
		The colors indicate each node's community. 
		Black nodes and dashed edges indicates, respectively, removed nodes and edges. The numbers indicate order in which the nodes are removed.
		\label{fig:dyn_example_graph}}
\end{figure}

The original graph before reduction is displayed in the left panel of Figure \ref{fig:dyn_example_graph}. 
Each node is labeled either by the color green or red. 
The graph is constructed from three subgraphs {\bf a}, {\bf b} and {\bf c}, that each has 20 nodes and one of the two labels:
\begin{description}
	\item[Graph a:] Average node degree $d = 3$, ratio of
	between-block edges over within-block edges $\beta=0.2$
	\item[Graph b:] Average node degree $d = 2$, ratio of
	between-block edges over within-block edges $\beta=0.4$
	\item[Graph c:]  Average node degree $d = 2$, ratio of
	between-block edges over within-block edges $\beta=0.2$, and an
	unbalanced community proportion; a probability of 3/4 for community 1 (red)
	and a probability of 1/4 for community 2 (green)
\end{description}
To link graphs a, b and c, some random edges are added between nodes from
the same community\footnote{For each pair of nodes between a pair of graphs, say Graph {\bf a} and
	{\bf c}, a new link is randomly sampled with a probability of 0.01, and
	links  connecting two nodes from the same community are kept.}. 
Nodes without any edges are also removed. 
This gives us a a total of 54 nodes, where 23 are red and 31 green. 
The task is to find an embedding for this graph and examine how well this embedding is able to distinguish between the two types of nodes. Further, the aim is to study this embedding and the classification scores as a function of time, as nodes (and their associated edges) are removed randomly. Finally, how does the classical PCA compare with the nonlinear embeddings in this process?

First, the network is embedded in $\mathbb{R}^m$ with $m=64$ using the Skip-Gram routine node2vec with (cf. section 5.3.2 and 6.5 of TJL) $L=30$ nodes in each random walk and $\gamma = 200$ walks per node, and a word2vec window length of $K=10$, where all nodes are included. The second step is to reduce the point cloud in $\mathbb{R}^{64}$ to $\mathbb{R}^2$, i.e., the visualization step using PCA and the two visualization algorithms t-SNE and UMAP with two parameter choices for t-SNE and three parameter choices for UMAP. More details of these embedding/visualization routines are given in sections 6.1 and 6.3 of TJL (the LargeVis algorithm is omitted here to obtain a simpler presentation.)

The original graph before reduction is displayed in the left panel of Figure \ref{fig:dyn_example_graph}, while the right panel shows the last stage (20th stage) where 19 nodes have been randomly removed. The removed nodes are numbered according to the iteration in which they are removed. In the right panel, removed nodes have been marked by black dots and the associated removed edges by dashed lines. The removal of nodes eventually resulted in a non-connected graph.

\begin{figure}
	\begin{center}
		\includegraphics[width=0.9\textwidth]{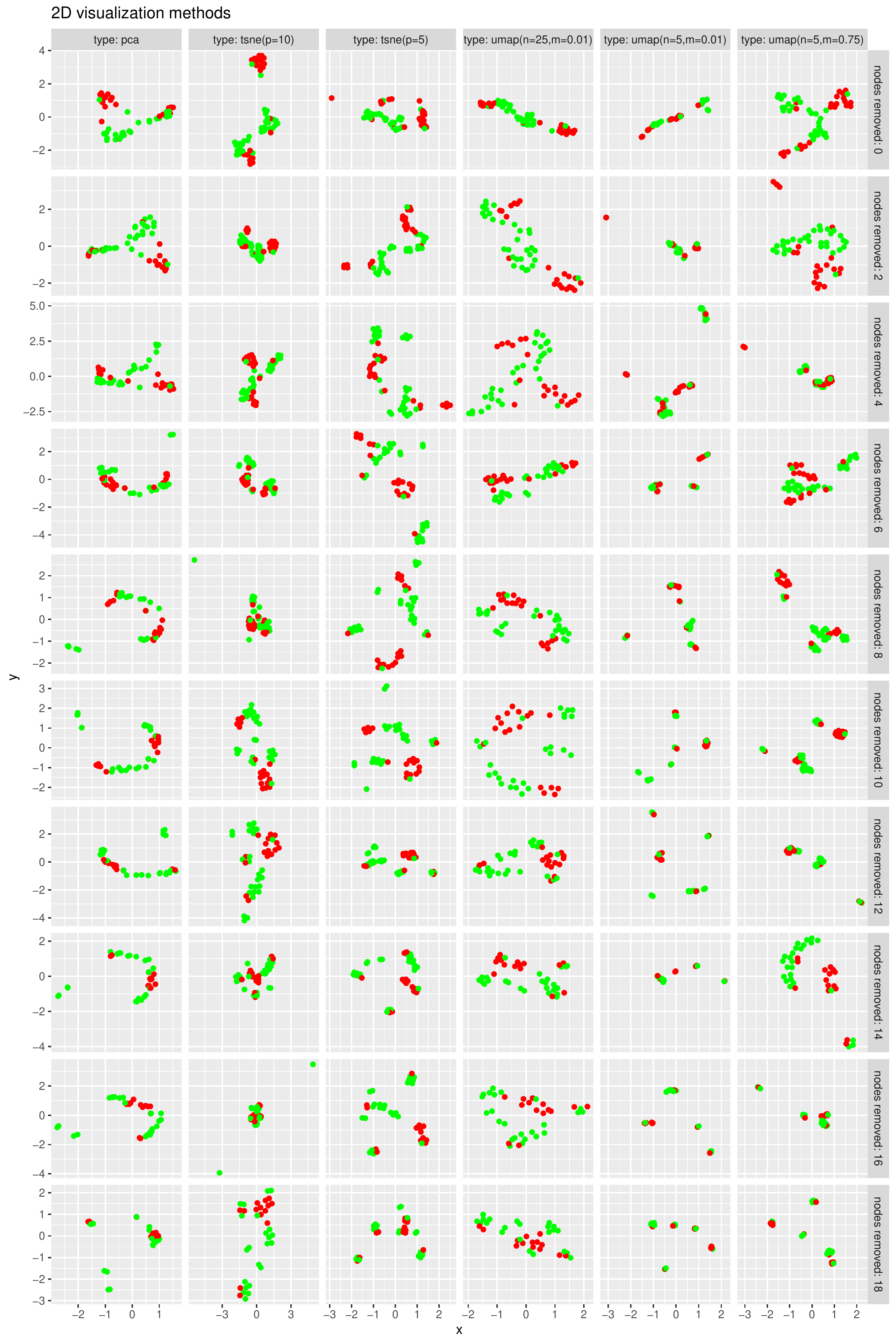}  
	\end{center}
	\caption{Visualizations of every second stage of the node2vec embeddings in the dynamic embedding illustration.
		\label{fig:dyn_example_vis}}
\end{figure}

Figure \ref{fig:dyn_example_vis} gives the embedding structure for every second stage in the iteration process, where the x- and y-axis of the visualization approaches have been standardized (mean zero and standard deviation 1) to ease comparison.
The uppermost row of this figure displays the embeddings for the complete graph in Figure \ref{fig:dyn_example_graph}. 
It is seen that the group structure of Figure \ref{fig:dyn_example_graph} is well taken care for most methods, but the shape of the embedding (as expected) vary considerably from one method to another. The problem mentioned in the second paragraph of this section is obvious in Figure \ref{fig:dyn_example_vis}. See for example the change in embeddings for PCA where the locations of red and green dots move around quite a bit as  nodes are removed. Despite this behavior, the internal community structure does not change that much. Such a lack of directional invariance appears to a smaller or lesser degree for all of the methods. Separate experiments also demonstrates that such directional changes can occur when different seeds are use for each iteration. Clearly one has to be aware of this fact to avoid misinterpretation of visualization plots.

\begin{figure}[ht!]
	\begin{center}
		\includegraphics[width=\textwidth]{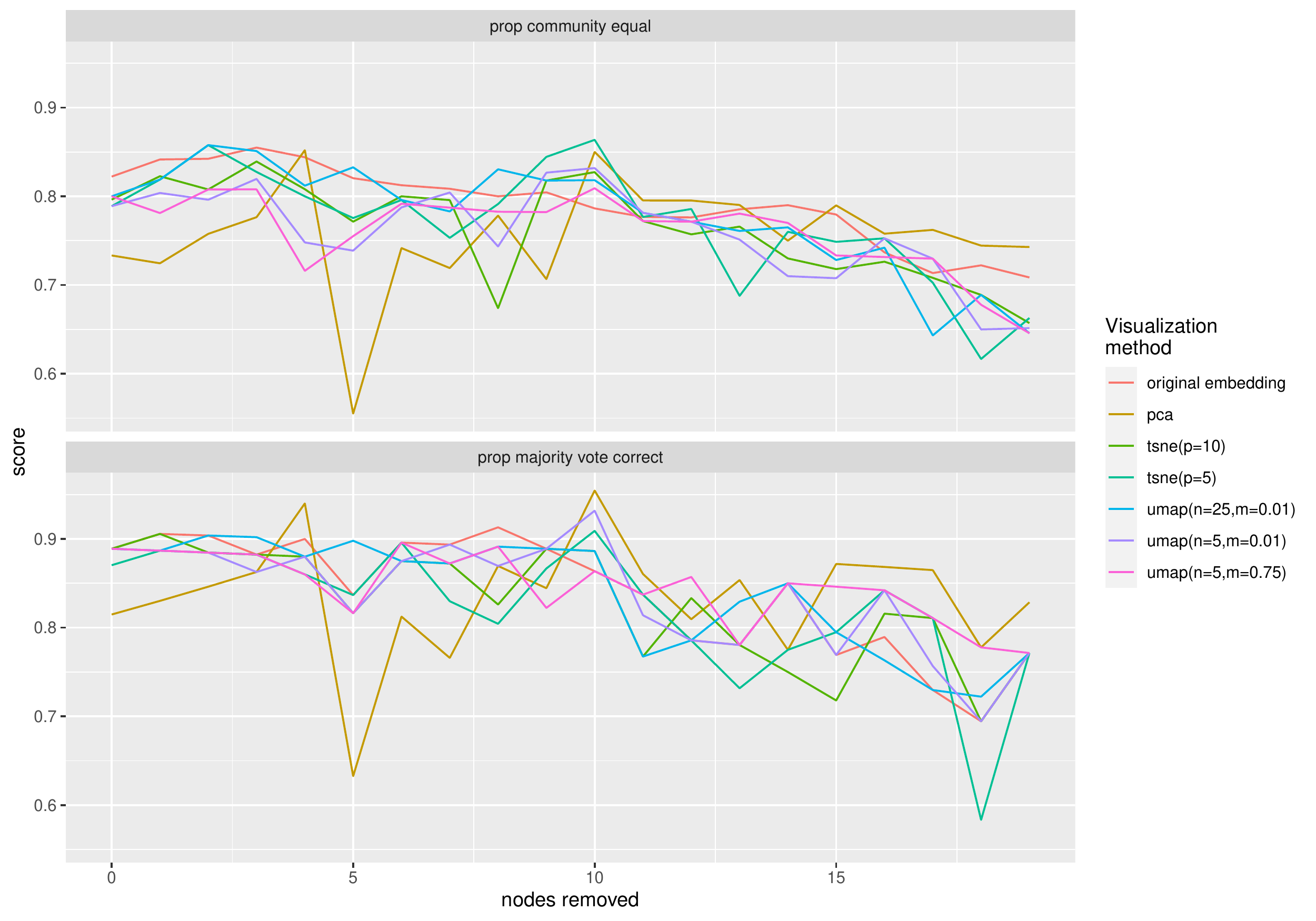}  
	\end{center}
	\caption{Classification scores plotted as a function of stage/node removal in the dynamic embedding illustration. The classification scores are based on a $k$-nearest neighbors algorithm with $k=5$.
		\label{fig:dyn_example_scores}}
\end{figure}

If the relative structure of groups of red and green dots in the sequence of embeddings is more or less invariant, then the hope is that the classification scores can still be meaningful and relatively stable from one iteration to another. This is confirmed by Figure \ref{fig:dyn_example_scores}. There the classification scores are plotted as a function of time for two basic classification algorithms based on k-nearest neighbors. 
The upper panel of Figure \ref{fig:dyn_example_scores} shows the classification scores when the class of a node is determined using the average of the 5 nearest neighbors; in the lower panel of the figure, the class is determined by the majority vote among the 5 nearest neighbors. The curve identified by ``original embedding'' gives the classification results for the 64-dimensional Skip-Gram embedding in step 1. 

Both figures indicate that PCA are clearly inferior to the others in the initial stages, but does well for higher iterations. The scores are quite stable as a function of time except for a possible jump at the removal of the 11th node after which the classification scores steadily decrease. The 11th node can be identified in Figure \ref{fig:dyn_example_graph} and it is presumably strategically located in the upper group of the red nodes. The main pattern is intuitively reasonable and it seems to be largely independent of any changes in orientation of the embeddings. The low score of the PCA in the initial stages is apparent also from the embeddings plot in Figure \ref{fig:dyn_example_vis}. In particular, see the embedding with 6 nodes removed, which mirrors the low classification scores for PCA for iteration 6. In both Figure \ref{fig:dyn_example_scores}a and \ref{fig:dyn_example_scores}b, PCA has a very low score at iteration 5. Inspecting the corresponding embedding plot (iteration 5 not shown in Figure \ref{fig:dyn_example_vis}) it is seen that this is not due to directional instability but simply that PCA at this iteration produces an embedding with much overlap, which in turn makes it harder to classify. To some degree this is present also for iteration 6.

Clearly one cannot draw general conclusions based on this special example, but the example does illustrate some of the problems that one can be expected to encounter in a more general dynamic situation. There is a need both for more experiments and for more theory.

\subsection{Statistical modeling of dynamic networks}

What is perhaps the largest difference between the dynamic systems we have been treating in this paper and an ordinary time series dynamic situation is the absence in dynamic networks of an explicit recursive system like the one that is present in the autoregressive model (possibly nonlinear and multivariate). Recently there have been some models tending in this direction (see also TJL).

A recent example of rigorous statistical modeling of a dynamic network is \cite{zhu:pan:li:liu:wang:2017}. They model the network structure by a network vector autoregressive model. This model assumes that the response of each node at a given time point is a linear combination of (a) its previous value, (b) the average of connected neighbors, (c) a set of node-specific covariates and (d) independent noise. More precisely, if $n$ is the number of nodes, let $Y_{it}$ be the response collected from the $i$th subject (node) at time $t$. Further, assume that a $q$ dimensional node-specific random vector $Z_i = (Z_{i1},\ldots,Z_{iq})^{T} \in \mathbb{R}^{q}$ can be observed. Then the model for $Y_{it}$ is given by
\begin{equation}
Y_{it} = \beta_0+Z_i^{T} \gamma + \beta_1 n_i^{-1} \sum_{j=1}^{n} a_{ij}Y_{j,t-1} + \beta_2 Y_{i,t-1} + \varepsilon_{it}.
\label{5.1}
\end{equation}
Here, $n_i = \sum_{j \neq i} a_{ij}$, $a_{ii}=0$, is the total number of neighbors of the node $v_i$ associated with $Y_i$, so it is the degree of $v_i$. The term $\beta_0+Z_i^{T} \gamma$ is the impact of covariates on node $v_i$, whereas $n_i^{-1} \sum_{j=1}^{n} a_{ij}Y_{j,t-1}$ is the average impact from the neighbors of $v_i$. The term $\beta_2 Y_{i,t-1}$ is the standard autoregressive impact. Finally the error terms $\{\varepsilon_{it}\}$ are assumed to be independent of the covariates and iid normally distributed.

Given this framework, conditions for stationarity are obtained, and least squares estimates of parameters are derived and their asymptotic distribution found. 

The authors give an example analyzing a Sina Weibo data set, which is the largest twitter-like social medium in China. The data set contains weekly observations of $n=2{,}982$ active followers of an official Weibo account. An extension of the model \eqref{5.1} is contained in \cite{zhu:pan:2020}. 

There are a number of differences between the network vector
autoregression modeled by equation \eqref{5.1} and the dynamic network
embeddings treated earlier in this section. First of all, \eqref{5.1}
treats the dynamics of the nodes themselves and not of an
embedding. Even if the autoregressive model do introduce some
(stationary) dynamics in time, the parameters are static; i.e.\ no new
nodes are allowed, and the relationship between them is also static as
modeled by the matrix ${\bf A} = \{a_{ij}\}$. From this point of view, as the authors are fully aware of, the model \eqref{5.1} is not realistic for the dynamics that takes place in practice for many networks.
On the other hand, the introduction of a stochastic model that can be analyzed by traditional methods of inference is to be lauded. A worthwhile next step is to try to combine more realistic models with a stochastic structure (e.g.\ regime type models for the parameters as in \cite{ludk:eckl:neal:2018} in the  context of the dcSBM model). The hope is that it will be amenable to statistical inference. \cite{kram:2019} treats dynamic networks with a fixed number of nodes, but where the dynamic structure is modeled by a doubly stochastic matrix.

For some very recent contributions to network autoregression, see \cite{armi:foki:krik:2022} and references therein. \cite{armi:foki:2022a} consider integer valued network variables and analyze linear and log linear Poisson autoregressive networks. This is motivated by the fact that many net variables take discrete values. In \cite{armi:foki:2022b} nonlinear models and tests for linearity are introduced.

\subsection{Concluding remarks}
We have given a survey of embeddings of time series and dynamic networks. 
We have covered dynamic factors for time series, and dynamic versions of nonlinear embeddings, topological data analysis embeddings, and  network embeddings. 
The embeddings have been illustrated  by two groups of simulated examples. 
The differences between a more or less purely algorithmic approach and an approach based on more statistical modeling have been pointed out, and we have seen that an algorithmic approach is clearly dominating. 
The literature on dynamic embeddings is much more sparse than the static case. 
This holds even though the dynamic embeddings could be far more realistic for many practical cases.
Throughout the review, we have pointed out a number of open modeling problems. 
We encourage time series analysts in particular, but also more general mathematical statisticians, to try to deal with these.

\subsection*{Acknowledgments}
This work was supported by the Norwegian Research Council grant 237718 (BigInsight).

\bibliography{dagemb}



\section*{Supplementary material}

# Supplement to "Some recent trends in embeddings of time series and dynamic networks"

Dag Tjøstheim[1,2]    Martin Jullum[2]    Anders Løland[2]

[1]University of Bergen
[2]Norwegian Computing Center

## 1   Deep learning networks

Neural networks are used for a number of problems in prediction, classification and clustering. Currently, there is an intense activity involving among other things deep learning, where some remarkable results have been obtained. See Schmidhuber (2015) for an overview that is still very useful (even though it was written seven years ago).

In this section we explain the principle for a single layer artificial network and then go on to multiple layer networks and deep learning. Assume that we are given an $n$-vector $x$ as input which may be thought of as a sliding window of data. In a neural network approach one is interested in transforming $x$ via linear combinations of its components and possibly a nonlinear transformation of these linear combinations. Those transformations constitute what is called a hidden layer.

Given the input layer, the first step in forming the hidden layer is to form linear combinations

$$h_i = \sum_{j=1}^{n} w_{ij} x_j + w_{i0}, \tag{1}$$

where $i = 1, \ldots, m$, and where $b_i = w_{i0}$ is a so-called bias term.

In the case of one hidden layer, the output layer is given in terms of the hidden units as

$$y_j = \sum_{i=1}^{m} w'_{ij} h_i, \tag{2}$$

for $j = 1, \ldots, q$, where $q$ in general may be different from $n$.

In a classification problem, $y_j$ may be associated with an unnormalized probability for a class $j$, which is related to an appropriate neighborhood of a node $v_j$ in a network. In such cases the output layer in (2) is also transformed. A common transformation is the so-called softmax function given by

$$\text{softmax}(y_j) = \frac{\exp(y_j)}{\sum_{i=1}^{q} \exp(y_i)}. \tag{3}$$

This is recognized (if there is no hidden layer) as the multinomial logistic regression model which is a standard tool in classification.

In a $s$-step time series forecasting problem, using the notation $Y_t = y_j$, $Y_t$ may successively be $X_{t+s}$ for an input training sets $\{X_1, \ldots, X_t\}$.

Using a training set, the coefficients (or weights) $w_{ij}$ and $w'_{ij}$ are determined by a penalty function measuring the distance between the output $y$ and the target vector $t$, for example measured by the loss function $E = ||y-t||^2$. In a classification and clustering problem the training set consists of input vectors $x$ belonging to known classes $i$. The target vector is a so-called "one hot" vector having 1 at the component $j$ for the given target and zeros elsewhere. The weights are adjusted such that the output vector is as close as possible to this vector. The same, but a simpler procedure due to a simpler output, is followed for forecasting.

The error function is evaluated for each of the samples coming in as inputs, and the gradient of the error function with respect to $y$ is evaluated with the weights being re-computed and updated in the direction of the gradient by stochastic gradient descent.

The weights $w'_{ij}$ for the output layer are computed first. Then, $w_{ij}$ are adjusted via the chain differentiation rule using so-called back propagation. Details are given in e.g. the appendix of Rong (2016). Schematically this may be represented by

$$w_{ij}^{(\text{new})} = w_{ij}^{(\text{old})} - \varepsilon \frac{\partial E}{\partial w_{ij}},$$

where $\varepsilon$ is a sensitivity parameter. Initial values for the weights can be chosen by drawing from a set of uniform variables.

This simple one-layer forward scheme has been used with considerable success (Mikolov et al., 2013a,b) in graph embedding as described in several sections of TJL. Dynamic embedding of graphs is described specifically in Section 5 of TJL.

However, it has turned out that in many problems a much improved performance can be obtained by deep multiple layer networks, whose learning is said to be done by deep learning.

For iid data mainly treated in Section 3 of TJL, the relevant problem is the clustering and discrimination problem, the forecasting problem being meaningless. For the time series case treated in this paper a main emphasis is on prediction, and the following will be phrased with time series notation $\{X_t\}$. A prime advantage with deep learning is then that hidden units can be connected to each other recursively in time. Using the framework and notation of Lim and Zohren (2021), the goal is to predict future values of a target $Y_{i,t}$, for a given entity $i$ at time $t$. Each entity can represent a logical grouping of temporal information – such as measurements from individual weather stations in climatology, or vital signs from different patients in medicine, a selection of economic time series – and can be observed at the same time, the simplest one-step-ahead forecasting models take the form

$$\hat{Y}_{i,t+1} = f(Y_{i,t-k:t}, X_{i,t-k:t,s_i}),$$

where $\hat{Y}_{i,t+1}$ is the model forecast, $Y_{i,t-k:t} = \{Y_{i,t-k}, \ldots, Y_{i,t}\}$, $X_{i,t-k:t} = \{X_{i,t-k}, \ldots, X_{i,t}\}$ are observations of the target and exogenous inputs respectively over a look-back window of length $k$, and $s_i$ are static meta-data associated with the entity (e.g. station location in a meteorology network). The function $f$ is the prediction function learned by the model. It contains the weight functions, the hidden units of the hidden layers, which can be chosen in various ways as shown in the following. The model is phrased here in terms of univariate forecasting, but can be extended to multivariate forecasting without loss of generality (Sen et al., 2019; Wen et al., 2017; Li et al., 2018; Salinas et al., 2019a).

## 1.1 Convolution neural networks (CNN)

There are several ways of choosing the function $f$. Convolution neural networks (CNN), Gu et al. (2018), were originally designed for spatial data but have been adapted to time series where researchers use multiple layers of causal convolutions. For an intermediate layer $l$, each causal convolution filter, in terms of hidden units, takes the form

$$h_t^{l+1} = A\Big((\mathbf{w} * h)(l, t)\Big) \qquad (4)$$

with the convolution between weights and hidden units given by

$$(\mathbf{w} * h)(l, t) = \sum_{\tau=0}^{k} \mathbf{w}(l, \tau) h_{t-\tau},$$

In (4), $h_t^l$ is an intermediate state at layer $l$, and $\mathbf{w}$ is a fixed matrix of convolution weights at layer $l$. The function $A$ is an activation function such as a sigmoidal function.

The fixed filter weights, not depending on time $t$, is a feature inherited from the spatial case. In addition, CNNs are only able to use inputs within its fixed look-back window to make forecasts. It is necessary with a carefully tuned choice of $k$ to take advantage of all relevant historical information. It is worth noting that a single causal CNN layer with a linear activation model is equivalent to an autoregressive model.

## 1.2 Recurrent neural networks (RNN)

Recurrent neural networks (RNN) do not require an explicit fixed specification of a look-back window as is the case for CNN. Given the natural interpretation of time series prediction as sequences of inputs and targets, many forecasting applications have been developed for temporal forecasting applications (Salinas et al., 2019b; Rangupuram et al., 2018; Lim et al., 2020). It has also been used in natural language processing (Young et al., 2018). At its core, RNN cell units contain an internal memory state which acts as a compact memory of past information. The memory state is recursively updated from new observations at each time step.

Recurrent neural networks are originally based on work by Rumelhart et al. (1986). Since then, a number of different RNN architectures have been developed. Just to give a flavor of these networks, we present the model equations for the three layer Elman network. This network is essentially a three layer network, with an input layer $\{X_t\}$, a hidden layer $\{h_t\}$ and an output layer $\{Y_t\}$. At each time step the input is fed forward and a learning rule is applied using a training set. The fixed back-connections save a copy of the previous values of the hidden units in the context unit (since they propagate over the connections before the learning rule is applied). Thus the network can maintain a sort of state variable allowing it to perform such tasks as sequence-prediction that are beyond the power of a standard multilayer perceptron. The model equations for the Elman network are

$$h_t = \sigma_h(\mathbf{W}_h X_t + \mathbf{U}_h h_{t-1} + b_h),$$
$$Y_t = \sigma_y(\mathbf{W}_y h_t + b_y),$$

where $X_t$ is an input vector, $h_t$ a hidden layer vector, $Y_t$ an output vector, $\mathbf{W}, \mathbf{U}, b$ are parameter matrices and a parameter vector $b$, $\sigma_h$ and $\sigma_y$ are activation functions, e.g.

of sigmoidal form. These should be compared to the simple feedforward system (1) and (2).

Jordan networks are similar to Elman networks. The context units are fed from the output layer instead of from the hidden layer, so that the model equations are

$$h_t = \sigma_h(\mathbf{W}_h X_t + \mathbf{U}_h Y_{t-1} + b_h),$$

$$Y_t = \sigma_y(\mathbf{W}_y h_t + b_y).$$

Elman and Jordan networks are also known as simple recurrent networks (SRN). There exist considerably more complicated recurrent networks with a number of layers as in the CNN network, but the idea of recursion in time is kept.

## 1.3 Long-short term memory (LSTM) recurrent networks

The long term gradients which are back-propagated can tend to zero or explode, i.e. tend to infinity, in classic RNNs. The problem is computational (or practical) when training the network using back propagation. The so-called long-short term memory (LSTM) recurrent network is designed to counter this problem. LSTM units partially solve the vanishing gradient problem because LSTM units allow gradient also to flow unchanged. However, LSTM networks can still suffer from the exploding gradient problem.

The vanishing gradient was first analyzed in a diploma thesis by Sepp Hochreiter at the Technological University of Munich under the guidance of Jürgen Schmidhuber. After having been published in a technical report as long short term memory and as a conference proceedings, the full LSTM paper appeared in 1997 in Neural Computation (Hochreiter and Schmidhuber, 1997). Since then there have been a substantial theoretical and not the least applied advances of the method. Now the original LSTM paper stands with a hefty more than 67 000 citations. The paper has become the most cited neural network article of the 20th century, and has been applied with considerable success to topics such as unsegmented, connected handwriting recognition, speech recognition, machine translation, robot control, video games and health care. LSTM is particularly well suited to classifying, and to processing and making predictions of time series data, since there can be lags of unknown duration between important events in a time series.

A common LSTM unit is composed of a cell unit, an input gate, an output gate and a forget gate. The cell remembers values over arbitrary time intervals, and the three gates regulate the flow of information into and out of the cell.

## 2 Simplical complexes

First, recall the definition of a simplex: Given a set $\mathbb{X} = \{X_0, \ldots, X_k\} \subset \mathbb{R}^p$ of $k+1$ "affinely independent vectors" (i.e., the vectors $(X_0, X_1, \ldots X_k)$ are linearly independent), the $k$-dimensional simplex $\sigma = [X_0, \ldots, X_k]$ spanned by $\mathbb{X}$ is the convex hull of $\mathbb{X}$. For instance, for $k = 1$ the simplex is simply given by the line from $X_0$ to $X_1$. The points of $\mathbb{X}$ are called the nodes of $\sigma$ and the simplices spanned by the subsets of $\mathbb{X}$ are called the faces of $\sigma$. A geometric simplical complex $K$ in $\mathbb{R}^p$ is a collection of simplices such that (i) any face of a simplex of $K$ is a simplex of $K$, (ii) the intersection of any two simplices of $K$ is either empty or a common face of both.

For a point cloud a graph can be constructed by connecting points that satisfy a neighborhood criterion, e.g. the distance is under a certain threshold. See for instance

Sections 3.4 and 5.1 in TJL. This leads to the standard notion of a neighboring graph from which the connectivity of the data can be analyzed and clustering can be obtained, including non-convex situations, as described by the Ranunculoid example. Using simplical complexes, where simplical complexes of dimension 1 are graphs, one can go beyond this simple form of connectivity. In fact a central idea in TDA (topological data analysis) is to build higher dimensional equivalents of neighboring graphs by not only connecting pairs but also $(k + 1)$-tuples of nearby data points. This enables one to identify new topological features such as cycles and voids and their higher dimensional counterparts. Regarding embedding of networks, as treated in Section 5, such a technique could possibly be used to discover cycles in networks such as criminal rings in fraud detection, say.

Simplical complexes are mathematical objects that have both topological and algebraic properties. This makes them especially useful for TDA. There are two main examples of complexes in use. They are the Vietoris-Rips complex and the Čech complex. The Vietoris-Rips complex $V_\varepsilon(\mathbb{X})$ can be introduced in a metric space $(M, d)$. It is the set of simplices $\mathbb{X} = [X_0, \ldots, X_k]$ such that $d_{\mathbb{X}}(X_i, X_j) \leq \varepsilon$ for all $(i, j)$.

These definitions should be compared to the use of ball-coverings in equation (7) of the main paper. The homology of $V_\varepsilon$ can be computed using basic matrix operations. All relevant computations can be reduced to linear algebra. This gives a method of computing homology and persistent homology relating the complexes as $\varepsilon$ varies as briefly mentioned in our simple introductory example of chain of circles, or the more involved example involving Ranunculoids.

Given a subset $\mathbb{X}$ of a compact metric space $(M, d)$, the families of Vietoris-Rips complexes, $\{V_\varepsilon(\mathbb{X})\}_{\varepsilon \in \mathbb{R}}$ are filtrations, that is, nested families of complexes. As indicated earlier, the parameter $\varepsilon$ can be considered as a data resolution level at which one considers the data set $\mathbb{X}$.

As in the introductory example in Section 4 of the main paper, the homology of a filtration $\{F_\varepsilon\}$ changes as $\varepsilon$ increases: new connected components can appear, existing components can merge, and loops and cavities may appear or be filled. Persistence homology tracks these changes, identifies the appearing features, and attaches a lifetime to them in the persistence diagram.



\end{document}